\documentclass[sigplan,screen]{acmart}

\usepackage{graphicx} 
\usepackage{soul} 
\usepackage{cleveref}

\title{Generate Compilers from Hardware Models!}

\author{Gus Henry Smith}
\affiliation{%
    \institution{University of Washington}
    \city{Seattle}
    \country{USA}}
\email{gussmith@cs.washington.edu}

\author{Ben Kushigian}
\affiliation{%
    \institution{University of Washington}
    \city{Seattle}
    \country{USA}}
\email{benku@cs.washington.edu}

\author{Vishal Canumalla}
\affiliation{%
    \institution{University of Washington}
    \city{Seattle}
    \country{USA}}
\email{vishalc@cs.washington.edu}

\author{Andrew Cheung}
\affiliation{%
    \institution{University of Washington}
    \city{Seattle}
    \country{USA}}
\email{acheung8@cs.washington.edu}

\author{Ren{\'e} Just}
\affiliation{%
    \institution{University of Washington}
    \city{Seattle}
    \country{USA}}
\email{rjust@cs.washington.edu}

\author{Zachary Tatlock}
\affiliation{%
    \institution{University of Washington}
    \city{Seattle}
    \country{USA}}
\email{ztatlock@cs.washington.edu}

\date{April 2023}

\begin{abstract}
Compiler backends\footnote{
  We broadly define
    a \textit{compiler backend}
    as any program
    that modifies,
    optimizes,
    or lowers 
    high-level, 
    hardware-independent code
    into low-level,
    hardware-specific code.
  This broad definition
    includes
    software compilers 
    like \texttt{gcc}
    and libraries
    like CUDA,
    but also hardware compilers
    like FPGA synthesis
    or High-Level Synthesis (HLS)
    tools.
}
  should
  be automatically generated
  from
  hardware design language (HDL)
  models of the hardware they target.
Generating compiler components
  directly from HDL
  can provide stronger correctness guarantees,
  ease development effort,
  and encourage hardware exploration.
Past work has already
  championed this idea;
  here we argue that advances in
  program synthesis
  make the approach more feasible.
We present a concrete example
  by demonstrating how FPGA
  technology mappers
  can be automatically generated
  from SystemVerilog models
  of an FPGA's primitives
  using 
  program synthesis.
\end{abstract}

\begin{document}

\maketitle

\begin{figure}
    \centering
    \includegraphics[width=200pt]{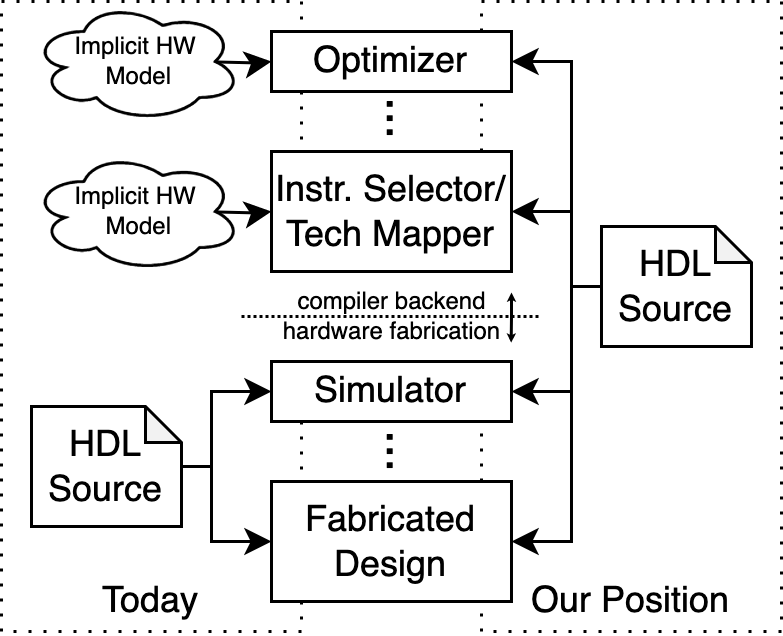}
    \caption{
How the components of 
  a software/hardware toolchain
  for a piece of hardware
  are built today:
  current state of the world (left)
  vs.~according to our position (right).
  }
    \label{fig:firstpage}
\end{figure}

\section{Our Position}

The semantics of HDLs
  are very rich.
From the advanced
  type systems of 
  new
  languages
  such as Aetherling~\cite{durst2020type}
  or Filament~\cite{nigam2023modular},
  to the high-level, algorithmic
  expressiveness
  of High Level Synthesis,
  hardware design languages
  convey much useful information
  about the hardware they describe.
Even stalwart SystemVerilog and VHDL
  accurately capture 
  a precise description
  of how a hardware design functions,
  including the ability
  to specify low-level details
  like latency.

Despite its richness,
  the HDL description
  of a hardware design
  is currently
  only used by the lowest layers
  of the software/hardware toolchain.
\Cref{fig:firstpage} (left)
  visualizes this.
The HDL model of a design
  \textit{is} used to build
  simulators and
  compile
  the final fabricated design,
  but the same model
  \textit{is not} used 
  when building higher-level
  toolchain components
  such as code optimizers
  or instruction selectors.
Instead,
  these parts of the toolchain
  often contain handwritten
  (and sometimes implicit)
  models of the target hardware,
  e.g.,
  a model of the hardware's memory
  hierarchy
  built into the optimizer.

It is our position
  that 
  \textbf{compiler backends
  should be automatically generated
  from the HDL model
  of the target hardware.}
Automatically generating compiler backends
  \textbf{(1) provides stronger 
    correctness guarantees}
  as compiler components no longer rely
  on handwritten,
  implicit,
  potentially buggy models of hardware.
Instead, compiler backends would rely
  on the same HDL source
  from which the hardware is fabricated,
  guaranteeing
  that the compiler's hardware model
  matches the fabricated hardware.
For similar reasons,
  automatically generating
  compiler backends
  \textbf{(2) 
    reduces compiler development effort}
  by removing the need
  to build a duplicate hardware model
  into the compiler backend.
Lastly, we believe
  this approach
  \textbf{(3) encourages hardware exploration.}
By providing more confidence
  in the correctness of the toolchain
  and reducing the burden of building
  a compiler for a new piece of hardware,
  hardware designers will be emboldened
  to experiment with new designs.

Furthermore, it is our position that
  \textbf{recent advances
  in programming languages
  make automated compiler construction
  feasible.}
The idea of automatically generating
  compiler backends is not new:
  previous work includes
  synthesizing instruction selectors%
  ~\cite{buchwald2018synthesizing,dias2010automatically,brandner2007compiler,daly2022synthesizing}
  and code generators~\cite{leupers1997retargetable,brandner2013automatic},
  among other work.
However, much of this work
  is a decade old, if not more,
  and does not benefit from
  advances in
  languages and type systems
  for hardware~\cite{durst2020type,nigam2023modular,nigam2020predictable},
  equational reasoning via equality saturation~\cite{tate2009equality,willsey2021egg},
  program synthesis~\cite{solar2008program,torlak2013growing},
  and machine learning for program generation~\cite{alon2019code2vec,austin2021program}.

To ground our position, we present
  a concrete example,
  in which we use SystemVerilog models
  of FPGA primitives
  to automatically build 
  technology mappers
  using modern program synthesis techniques.

\section{Generating Technology Mappers}

Technology mapping is an
  FPGA compiler backend step
  in which a high-level hardware design
  is lowered to use hardware 
  \textit{primitives}
  (small functional blocks)
  available on the target FPGA.
Currently, technology mappers
  are often implemented
  as hand-written pattern matchers,
  which look for patterns in high-level
  HDL code
  and rewrite them to instances
  of FPGA primitives.\footnote{
  For one example of these patterns in the open-source FPGA compilation tool Yosys~\cite{wolf2013yosys}, see \url{https://github.com/YosysHQ/yosys/blob/cee3cb31b98e3b67af3165969c8cfc0616c37e19/techlibs/xilinx/xcu_dsp_map.v}}
Some 
  automation does exist;
  the VTR project~\cite{rose2012vtr}
  seeks to automatically
  provide compiler backends
  for hardware
  given just an
  architecture description
  using tools like
  ABC~\cite{brayton2010abc}
  and
  ODIN-II~\cite{jamieson2010odin}.
  
Existing technology mapping
  approaches---%
  hand-written pattern matchers
  and automated tools---%
  have a number of weaknesses.
They fail to provide strong
  correctness guarantees:
  hand-written patterns can be
  incorrect.
They require significant
  developer effort:
  when an automated tool
  cannot support an FPGA primitive,
  developers must support
  the primitive by hand.
Finally, 
  current tools
  limit exploration:
  each new FPGA primitive
  represents a potentially high cost
  to support.

\begin{table}
\caption{FPGA primitives imported automatically (and thus available for technology mapping) from vendor-provided SystemVerilog models, with source lines of code  of the original SystemVerilog models.}
\centering
\normalsize
\label{table:imported-primitves}
\begin{tabular}{lrr}
 {\bf FPGA}   & \textbf{Primitive} & {\bf SystemVerilog} \\\hline
 Xilinx Ultrascale+  & LUT6 & 88       \\
                     & CARRY8 & 23       \\
                     & DSP48E2 & 1426       \\
 Lattice ECP5 & LUT2 & 5       \\
              & LUT4 & 7       \\
              & CCU2C & 60       \\
              & ALU24B & 672       \\
              & MULT18X18D & 985       \\
 SOFA~\cite{tang2021taping}      & frac\_lut4   &  69       \\
 Intel Cyclone & altmult\_accum& 1460 \\
\end{tabular}
\end{table}

We have prototyped
  a tool
  which 
  generates technology mappers
  automatically
  from the HDL models
  of the target FPGA.
Our tool
  automatically
  extracts bitvector semantics
  from the SystemVerilog models
  of FPGA primitives
  provided by each FPGA vendor.
We then apply
  \textit{program synthesis,}
  a technique which utilizes
  SMT solvers
  to generate programs.
We use the bitvector semantics
  extracted from each primitive's
  SystemVerilog model
  to check---%
  with the help of the solver---%
  whether the primitive
  can be configured
  to implement the
  input high-level hardware design.
Furthermore, we build an
  intermediate representation
  which allows for the construction
  of platform-independent
  \textit{templates,}
  which capture patterns
  common across FPGA architectures.

Our prototype approach
  to technology mapping
  provides strong correctness guarantees,
  reduces development effort,
  and can support hardware exploration.
Our approach's strong
  correctness guarantees
  come not only from our use of
  SMT solvers,
  but also from the fact
  that we use the primitive semantics
  extracted directly from SystemVerilog,
  rather than relying on
  handwritten, and potentially incorrect,
  semantics.
We quantify our approach's reduction
  of development effort
  by listing the primitives 
  automatically imported
  (and thus supported)
  by our tool in
  \cref{table:imported-primitves}.
Finally, our approach
  can encourage the exploration
  of new FPGA primitives,
  by quickly generating
  technology mappers
  for hardware prototypes
  during the development process.

\section{Conclusion and Future Directions}

We have argued that
  compiler backends should
  be automatically generated
  from the HDL models of the hardware
  they target.
Furthermore, we provided a 
  concrete demonstration
  of this idea
  via a prototype tool
  which generates FPGA
  technology mappers
  given the SystemVerilog models
  of an FPGA's hardware primitives.
  
We call on others in the field
  to revive this idea with us
  via the application
  of modern techniques,
  such as machine learning
  or equational reasoning via
  equality saturation.

\clearpage

\bibliographystyle{plainnat}
\bibliography{bib}
\end{document}